\documentstyle[prl,aps]{revtex}
\twocolumn
\begin{document}
\title{\large \bf Strongly Non-Equilibrium 
Bose-Einstein Condensation in a Trapped Gas}
\author {Boris Svistunov}
\address{
Russian Research Center ``Kurchatov Institute", 123182 Moscow, Russia}

\maketitle

\begin{abstract}
We present a qualitative (and quantitative, at the level of estimates) 
analysis of the ordering kinetics in a strongly non-equilibrium state 
of a weakly interacting Bose gas, trapped with an external potential.
At certain conditions, the ordering process is predicted to
be even more rich than in the homogeneous case. Like in the homogeneous 
case, the most characteristic feature of the full-scale non-equilibrium
process is the formation of superfluid turbulence.
\vspace{0.2cm} 
\vspace{0.5cm}
\end{abstract}
Kinetics of Bose-Einstein condensation (BEC) in a weakly interacting Bose gas
is one of the most fundamental problems of non-equilibrium statistical mechanics. 
The exciting progress in the experiment on BEC in ultracold gases initiated 
by the pioneer works \cite{BEC} opens up an opportunity of laboratory study of 
BEC kinetics. Though an important step in this direction has
been already made \cite{Miesner}, most non-trivial regimes of ordering kinetics have not 
been achieved yet.

In Refs.\ \onlinecite{Sv,KSS,KS1} (see also Ref.\ \onlinecite{ST}) a scenario of BEC in a 
strongly non-equilibrium self-evolving homogeneous gas was developed. 
The scenario involves a number of qualitatively different stages. 
The evolution starts with an explosion-like wave in the wavenumber space, 
propagating from higher energies towards the lower ones, and resulting in 
a formation of a specific power-law distribution of particles over the 
wavenumbers \cite{Sv}. Immediately after its 
formation, this distribution starts to relax, the relaxation having the form of a 
back wave in the wavenumber space \cite{Sv}. Simultaneously, in the 
low-energy region the so-called coherent regime sets in that leads to the formation 
of the quasicondensate state \cite{KSS}. Basically, the quasicondensate 
state corresponds to what is known in the theory of superfluidity as the state of 
superfluid turbulence. It can be viewed as a condensate containing a tangle of 
vortex lines
(plus a sharply non-equilibrium distribution of long-wave phonons) \cite{KS1}. 
The formation of the quasicondensate occurs very rapidly (characteristic time is
much smaller than the time of the wave formation) \cite{Sv,KSS}. In contrast to it,
the final stage of long-range ordering, associated with relaxing superfluid turbulence
and long-wave phonons, takes a macroscopically large (the size of the system
enters the answer) time \cite{KS1}.

From a common wisdom one may expect that at certain conditions 
strongly non-equilibrium BEC in a trapped gas should reproduce main 
qualitative features of the homogeneous
scenario. It is not surprising thus that the explosion-like 
initial kinetic stage \cite{Sv} is quite relevant to the case of the trapped 
gas \cite{Gardiner,Stoof} and is consistent with the experimental observations of 
Ref.\ \cite{Miesner}. Nevertheless, full-scale experimental realization of the scenario 
\cite{Sv,KSS,KS1} in a trapped gas is not guaranteed yet by just  
considerable deviation from the equilibrium in the initial state.
The point is that the situation with the BEC kinetics in 
ultracold gases is determined by a competition between large number of
particles and very small interaction. The latter circumstance leads to a rather 
large coherent length associated with the ordering process,
which for a trapped gas can turn out to be just larger than the size of the condensate,
in which case the character of the condensate formation will be drastically changed: 
Instead of quasicondensate state there will take place a build-up of 
a genuine condensate with
the initial form pre-determined by the external potential.
The theoretical analysis of the BEC kinetics in a trapped gas 
has been dealing mostly with this physical picture \cite{Gardiner,Stoof}, 
ignoring thus the nontrivial regime associated with the superfluid 
turbulence formation. 

The main goal of the present letter is to consider the superfluid-turbulence 
regime of BEC kinetics in an external potential.
We will demonstrate that the necessary 
and sufficient condition for this regime
to occur corresponds to violation of the Knudsen condition at a 
certain stage of self-evolution
within a certain region of coordinate space. While the external potential
together with the initial number and energy of particles determines the initial parameters
of the quasicondensate: the size $R_*$, the number of quasicondensate 
particles $\tilde{N}_0$, and typical initial separation between vortex 
lines in the vortex tangle, $r_0$, the very
process of the quasicondensate formation goes in a quasi-homogeneous {\it anti}-Knudsen 
regime when the external potential becomes {\it irrelevant} and, in principle, may be 
even turned off without any considerable effect on the process of the quasicondensate
formation.

In the experiments with trapped ultracold gases normally there takes place
the Knudsen regime, when the free path length of a particle with the energy $\varepsilon$,
$l_{\mathrm{free}}(\varepsilon, n_{\varepsilon})$ [which depends also on
occupation numbers $n_{\varepsilon}$ because of the stimulated scattering], 
is much larger than the typical radius of the 
particle's trajectory, $R_{\varepsilon}$. For definiteness we consider a parabolic
trap with all the three frequencies of the same order $\omega_0$. So that
$R_{\varepsilon} \sim \omega_0 v(\varepsilon)$ [$v(\varepsilon)$ is the
typical velocity corresponding to the energy $\varepsilon$], and the condition 
$l_{\mathrm{free}}(\varepsilon, n_{\varepsilon}) \gg R_{\varepsilon}$ is equivalent to
$\tau_{\mathrm{coll}}(\varepsilon, n_{\varepsilon}) \omega_0 \gg 1$, 
where ($\hbar=1$)
\begin{equation}
\tau_{\mathrm{coll}}(\varepsilon, n_{\varepsilon}) 
\sim 1 / m (a \varepsilon n_{\varepsilon})^2 
\label{tau_coll}
\end{equation}
is the collisional time. Here $a$ is the scattering length and 
$m$ is the particle mass.

Knudsen regime is a convenient starting point for analyzing kinetics 
in a potential. Almost in all qualitative aspects it corresponds to an
isotropic homogeneous case, since the distribution of particles depends
only on the two variables, $\varepsilon$ and $t$ (ergodic approximation). 
The scaling of the collision time remains the same, so that 
the main quantitative difference comes from the difference in the 
density of states. 

During some initial period of evolution the correlations between different 
single-particle modes are vanishingly small. Such a regime admits a description in
terms of kinetic equation. This stage is thus referred to as kinetic stage.
If the deviation from equilibrium is strong enough, the evolution at the kinetic 
stage results in the formation of a self-similar wave in the wavenumber space 
propagating in an explosion-like fashion from the high-energy region (where 
the particles are initially distributed) towards lower energy scales.
Corresponding solution of the kinetic equation has the form \cite{Sv}
\begin{eqnarray}
n_{\varepsilon}(t) = A \varepsilon_0^{-\alpha}(t) 
f(\varepsilon/\varepsilon_0(t)) \;, \;\;\; t \leq t_* \; , \label{wave_a}\\
\varepsilon_0(t) = B \mid t_* -t \mid^{1/2(\alpha-1)}\; . \label{wave_b}
\end{eqnarray}
Here $A$ and $B$ are dimensional constants depending on the initial condition
and related to each other by the formula 
$B=\mathrm{const} (m a^2 A^2)^{1/2(\alpha-1)}$.
The dimensionless function $f$ is 
defined up to an obvious scaling freedom.  
The explosion character of the evolution guarantees that the wave reaches
the point $\varepsilon=0$ at some finite time moment $t=t_*$ 
[$t=0$ corresponds to the beginning of evolution], the value of $t_*$ 
being on the order of the collisional time at $t=0$. 
Physically, the explosion-like evolution
is supported by the stimulation of the collision rate at the head of the wave,
$\varepsilon \sim \varepsilon_0$, by ever growing occupation numbers.
A remarkable feature about the solution (\ref{wave_a})-(\ref{wave_b})
is its universality: The coefficient $A$ is the {\it only} quantity related to a 
particular cooling regime or a particular form of the initial condition. 
Having in mind the most characteristic statement of the problem when
the self-evolution starts with an essentially non-equilibrium state of $N$ 
particles in a parabolic trap of typical frequency $\omega_0$, with a
typical occupation numbers $\gamma > 1$, we readily estimate $A$ by
extrapolating the solution (\ref{wave_a})-(\ref{wave_b}) to $t=0$:
\begin{equation}
A \sim \gamma^{1-\alpha /3} \omega_0^{\alpha}N^{\alpha /3} \; .
\label{A}
\end{equation}

Generally speaking, the index $\alpha$ in (\ref{wave_a})-(\ref{wave_b}) 
cannot be established from the scaling
properties of the collision term of the kinetic equation, 
being related thus to the particular
form of the latter \cite{note1}. It is possible, however, 
to specify lower and upper limits
for $\alpha$ following from the consistency of (\ref{wave_a})-(\ref{wave_b}) 
with the requirement that these formulae describe an explosion-like singularization
of distribution (rather than infinite-time shrinking). To this end we note that 
from the scale invariance of the collisional term it follows that $f(x)$ 
should behave at $x \gg 1$ like some power of $x$. Then, from the fact 
that at $t=t_*$ the occupation numbers are finite at
$\varepsilon > 0$ one concludes that
\begin{equation}
f(x) \to x^{-\alpha} \;\; \mathrm{at} \;\; x \to \infty \;  . 
\label{f}
\end{equation}
The requirement that the particle distribution does not shrink as a whole
implies that the number-of-particles integral for the distribution (\ref{wave_a}), 
(\ref{f}) is divergent at $\varepsilon \to \infty$. The number of particles
$N = \int w(\varepsilon) n_{\varepsilon} d \varepsilon \,$,
where the density of states $w(\varepsilon)$ is given by
\begin{equation}
w(\varepsilon) = \left\{ 
\begin{array}{ll}
V m^{3/2} \varepsilon^{1/2} / \sqrt{2} \pi^2 
\;\;\;  \mbox{(homogeneous)} \; , \\
2 \varepsilon^2 / \omega_x \omega_y \omega_z 
\;\;\;\;\;  \mbox{(parabolic potential)} \; .
\end{array} \right.
\label{w}
\end{equation}
Here $V$ is the volume of a homogeneous system and 
$\omega_x$, $\omega_y$, and $\omega_z$ are the frequencies 
of the 3D harmonic oscillator.
This immediately yields  $\alpha < 3/2$ for the homogeneous case
and $\alpha < 3$ for the case of parabolic potential. 
The condition $\alpha > 1$ is necessary 
for $\varepsilon_0(t)$ to approach zero at $t=t_*$. 
So we have $1 < \alpha < 3/2$ in the homogeneous case and
$1< \alpha < 3$ in the parabolic potential. The most accurate up-to-date 
numeric analysis of $\alpha$ in the homogeneous case was performed in Ref.\ \cite{ST} 
with the result $\alpha \approx 1.24$. In the case of 
parabolic potential $\alpha \approx 1.6$ \cite{Gardiner,Stoof}.

Formally, at $t=t_*$ the front of the wave $\varepsilon_0(t)$ reaches the energy
$\varepsilon=0$. However, somewhat earlier the solution (\ref{wave_a})-(\ref{wave_b})
becomes inapplicable in the region of very small energies.  In the homogeneous case
this occurs due to the change of the evolution regime (in corresponding low-energy
region) from weak turbulence to strong turbulence (the so-called coherent regime)
\cite{Sv,KSS}. In the external potential, however, the scenario is modified.
As we demonstrate below, what changes the solution (\ref{wave_a})-(\ref{wave_b})
at a certain stage of evolution of a trapped gas is either (i) discreteness
of the levels, or (ii) violation of the Knudsen regime in a certain region of
the coordinate space. Apart from a cross-over region, the two circumstances are
mutually exclusive: If discreteness of the levels becomes relevant, the Knudsen
regime will not be violated, and if the Knudsen regime is violated, the discreteness
of levels will not become relevant. Hence, the change of the evolution
in a trapped gas from the self-similar wave (\ref{wave_a})-(\ref{wave_b})
is governed by the following two parameters:
\begin{equation}
\xi = \tau_{\mathrm{coll}} \omega_0
 \;\;\; \mbox{(Knudsen parameter)} \;  , 
\label{xi}
\end{equation}
\begin{equation}
d = \tau_{\mathrm{coll}} / w(\varepsilon)
 \;\;\; \mbox{(discreteness parameter)} \;  . 
\label{d}
\end{equation}
Knudsen regime corresponds to $\xi \gg 1$; the discreteness of the
levels is irrelevant when $d \ll 1$. Note that $\tau_{\mathrm{coll}}$
essentially depends on $\varepsilon$ and $n_{\varepsilon}$, 
which ultimately means that both $\xi$ and $d$ are the functions of energy and time.

First, let us consider the case (i), where the condition $d \sim 1$ occurs when
the front of the wave $\varepsilon_0(t)$ reaches some energy $\varepsilon_d$.
As it follows from (\ref{tau_coll})-(\ref{wave_b}), (\ref{w}), the parameter $d$
at the front of the wave can be estimated as
\begin{equation}
d_{\varepsilon_0} \sim \omega_0^3 \varepsilon_0^{2(\alpha-2)} / 
m a^2 A^2  \;  . 
\label{d_front}
\end{equation}
From Eq.(\ref{d_front}) we see that $d_{\varepsilon_0}$ decreases with 
decreasing $\varepsilon_0$ only if $\alpha < 2$, which means that the case (i)
is possible only under this condition. [In this respect
the result $\alpha \approx 1.6$ \cite{Gardiner,Stoof} is of qualitative
importance!] Substituting $A$ from (\ref{A}) into (\ref{d_front}) and setting
$d_{\varepsilon_0} \sim 1$, we find
\begin{equation}
\varepsilon_d  \sim \omega_0 p^{-1/2(2-\alpha)}  \;  , 
\label{e_d}
\end{equation}
\begin{equation}
p = \omega_0 m a^2 \left( N^{\alpha/3} \gamma^{1-\alpha/3} \right)^2 \; .
\label{p}
\end{equation}
Here $p$ is the main dimensionless parameter which we will encounter several times.
By definition, in the case (i) the value of $\varepsilon_d$ Eq.(\ref{e_d})
should be greater than $\omega_0$, which implies $p \ll 1$. Below we will
see that $p \ll 1$ is not only a necessary, but also a sufficient condition
for the case (i) to occur: Under this condition the case (ii) is impossible.

To describe what happens after the front $\varepsilon_0(t)$
reaches the energy scale $\varepsilon_d$, we need to make an
assumption that the discreteness of the levels leads to a 
{\it slowing down} of the evolution cascade at 
$\varepsilon < \varepsilon_d$. Such an assumption \cite{note2} seems to be
quite natural in view of the fact that discrete harmonics of different 
frequencies are out of resonance, which should suppress the cascade
in the wavenumber space. If this is the case, the next stage of evolution
at the energies $\varepsilon < \varepsilon_d$ will be a fast equilibration
[with the characteristic time  
$\tau_{\mathrm{coll}}(\varepsilon = \varepsilon_d,
n_{\varepsilon}=A/\varepsilon_d^{\alpha})$; harmonics with energies
$\varepsilon \ll \varepsilon_d$ equilibrate due to the interaction
with harmonics $\varepsilon \sim \varepsilon_d$, rather than with each other].
Further evolution should lead to the destruction of the non-equilibrium
distribution $n_{\varepsilon}=A/\varepsilon^{\alpha}$ at 
$\varepsilon \gg \varepsilon_d$, the process having the form 
of a self-similar back cascade in the wavenumber space (back wave) \cite{Sv}:
\begin{equation}
n_{\varepsilon}(t) = A \varepsilon_0^{-\alpha}(t) 
\tilde{f}(\varepsilon/\varepsilon_0(t)) \;, \;\;\; 
\varepsilon > 0 \;, \;\;\; t > t_* 
\label{back}
\end{equation}
[$\tilde{f}(x) \to f(x) \;\;$ at $\;\; x \to \infty$], where $\varepsilon_0(t)$
is given by the same formula (\ref{wave_b}). Clearly, Eq.\ (\ref{back}) is
applicable only if $t-t_*$ greater than the above-mentioned time of 
equilibration of the harmonics $\varepsilon < \varepsilon_d$.
As follows from an estimate of corresponding terms of collision
integral \cite{Sv}, the back wave should create a quasi-equilibrium distribution 
at $\varepsilon \ll \varepsilon_0(t)$, that is $\tilde{f}(x) 
\propto 1/x$ at $x \ll 1$, with infinite at $t=t_*$ and ever decreasing 
afterwards temperature $\propto \varepsilon_0^{-\alpha}(t)$.

Since after the formation of the back wave all the harmonics with 
$\varepsilon \ll \varepsilon_0(t)$, including the lowest one (!),
are in a quasi-equilibrium, the process of long-range ordering in
the case (i) is somewhat trivial: The state of the harmonics
$\varepsilon \ll \varepsilon_0(t)$ is determined by the instant
temperature. In particular, the number of condensate
particles $N_0(t)$ follows from Eq.\ (\ref{back}) by conservation
of the total number of particles:
\begin{equation}
N_0(t) = {2A \over \omega_x \omega_y \omega_z} \, \varepsilon_0^{3-\alpha}(t)
\int_0^{\infty} d x \, x^2 [x^{-\alpha} - \tilde{f}(x)] \; .
\label{N_0}
\end{equation}
This universal law of the condensate build-up should take place until the
back wave cascade reaches the initial energy scale and the further
growth of $N_0$ becomes sensitive to the details of preparation of
the initial state.

We now turn to the most non-trivial case (ii), where the condition 
$\xi \sim 1$ occurs when the front of the wave $\varepsilon_0(t)$ reaches some 
energy $\varepsilon_{\xi}$. From (\ref{tau_coll})-(\ref{wave_b}) and
(\ref{w}) we have the following estimate for the parameter $\xi$ at the front of the wave:
\begin{equation}
\xi_{\varepsilon_0} \sim \omega_0 \varepsilon_0^{2(\alpha-1)} /
m a^2 A^2  \;  . 
\label{xi_front}
\end{equation}
Substituting $A$ from (\ref{A}) into (\ref{xi_front}) and setting
$\xi_{\varepsilon_0} \sim 1$, we find
\begin{equation}
\varepsilon_{\xi}  \sim \omega_0 p^{1/2(\alpha-1)}  \;  , 
\label{e_xi}
\end{equation}
which, in particular, means that the case (ii) can only occur if $p > 1$.
From Eqs.\ (\ref{e_d}) and (\ref{e_xi}) we conclude that, apart a cross-over
region (corresponding to $p \sim 1$), 
the following alternative takes place. Either discreteness of the 
low-lying levels becomes relevant ($p \ll 1$), or there takes place
a break-down of Knudsen regime ($p \gg 1$). Geometric size $R_*$ of the area
where the break-down of Knudsen regime takes place is obviously related to
$\varepsilon_{\xi}$ by $ m \omega_0^2 R_*^2 \sim \varepsilon_{\xi}$, which
gives $R_* \sim l_0 p^{1/4(\alpha-1)}$, where $l_0=1/\sqrt{m\omega_0}$ is the 
typical size of the single-particle groundstate wavefunction in the trap.

A natural question then is: What happens at distances $r < R_*$ from the center of 
the trap after the Knudsen regime is broken down? 
A self-consistent scenario of future 
evolution can be constructed by noticing that further
increase of the collisional rate due to the stimulated scattering naturally
leads to the {\it anti}-Knudsen regime, when the free-path-length of the
low-energy particles becomes much smaller than $R_*$. This immediately leads
to a quasi-homogeneous picture of evolution: the region
$r < R_*$ can be formally thought of as a number of independent homogeneous 
sells of the size much larger than the free-path-length, but much smaller 
than $R_*$. Evolution in each sell follows homogeneous scenario, ultimately 
leading to the formation of the quasicondensate. Indeed, in the anti-Knudsen
regime the evolution during the time period on the order of collision time
is insensitive to the external potential (the criteria for the Knudsen regime and for
the sensitivity to the potential within the collision time coincide). But this time is
enough to form the wave (\ref{wave_a})-(\ref{wave_b}) in the wavenumber space 
(with the exponent $\alpha$ now corresponding to the homogeneous case) 
and then to form quasicondensate. During the evolution in the wavenumber
space, the free-path length of the particles with $\varepsilon \sim \varepsilon_0(t)$
is getting progressively smaller, which renders the proposed scenario self-consistent.
  
A minor deviation from the pure homogeneous picture is that now the moment $t_*$
depends on the distance from the center of the potential, so that the coherent
regime first should start at $r=0$ (the point of maximal initial density) and then 
gradually occupy all the anti-Knudsen region up to $r \sim R_*$. Thus, the process
of quasicondensation takes on a form of wave propagating in the coordinate
space from $r=0$ towards $r \sim R_*$. When this wave reaches $r \sim R_*$,
the process of formation of the quasicondensate correlation properties is
finished, and within the region $r < R_*$ we have a state of superfluid
turbulence. 

Note that the fact that the quasicondensate formation in the anti-Knudsen
regime goes independently of the external potential should lead to certain
qualitative effects. For example, after the anti-Knudsen regime is well formed,
the trapping potential can, in principle, be {\it turned off} without any
dramatic effect on the (short-range) ordering process. Clearly, 
experimental realization of such a regime would be interesting and 
instructive from the fundamental point of view. Another effect, coming from
the same reason, is that the quasicondensate, just upon its formation,
is out of equilibrium with respect to a global coherent motion in the potential,
which means that a breathing mode for the quasicondensate motion 
should ultimately be excited in a trapped gas.

As far as the evolution in the anti-Knudsen region is of quasi-homogeneous
character, all the quantities characterizing this process can be
estimated from homogeneous relations of Refs.\ \cite{Sv,KSS,KS1}. As an
initial condition to the corresponding homogeneous case 
one should take an extrapolation of the inhomogeneous solution
(\ref{wave_a})-(\ref{wave_b}) to the region of breaking down the Knudsen regime.
Hence, the typical initial homogeneous single-particle energy and occupation
number are nothing else than $\varepsilon_{\xi}$ and  
$A/\varepsilon_{\xi}^{\alpha}$, respectively. Correspondingly, the number
of particles involved in the quasi-homogeneous BEC process can be estimated as
\begin{equation}
\tilde{N} \sim \gamma^{1-\alpha/3} N^{\alpha/3} p^{(3-\alpha)/2(\alpha-1)} \; .
\label{N_tilde}
\end{equation}
An important step of the qusi-homogeneous regime is associated with coming
the homogeneous back wave (\ref{wave_a})-(\ref{wave_b}) to the initial
energy scale (which in our case corresponds to $\varepsilon_{\xi}$).
At this stage the number of qussicondensate particles $\tilde{N}_0$ is on the order
of the total number of particles. Eq.(\ref{N_tilde}) thus yields an
estimate of the number of quasicondensate particles which will be 
created in the course of anti-Knudsen self-evolution in the system, even if
the trapping potential is turned off upon the formation of the 
anti-Knudsen regime.

The state of superfluid turbulence, which arises together with the
formation of the quasi-condensate local correlation properties, is characterized by
the typical separation between the vortex lines. This separation $r_0$ is minimal
just upon the quasicondensate formation, when $r_0$ is on the order
of the quasicondensate healing length. From the estimate of the quasicondensate
density upon its formation \cite{Sv} we find $r_0$ and  
come to the following relation:
\begin{equation}
R_*/r_0 \sim p^{\nu} \; , \;\;\;\; \nu = {4 \alpha_0 -3  \over 
4(\alpha-1)(2\alpha_0-1)} \; ,
\label{r_0}
\end{equation}
where $\alpha_0 \approx 1.24$ is the exponent $\alpha$ for the homogeneous
case. From (\ref{r_0}) we see once again that the formation
of the superfluid turbulence is possible only if $p \gg 1$.  

In conclusion, we note that in the experiment of Ref.\ \cite{Miesner}
the parameter $p$ is, roughly speaking, of order unity. This means
that the full-scale non-equilibrium BEC scenario cannot be seen with
these initial conditions. Theoretically, the most direct way
of achieving $p \gg 1$ is to radically increase $\omega_0$ 
immediately after the formation of the non-equilibrium initial state, 
during the time which is much less than the collisional time.

\end{document}